\documentclass[pra,superscriptaddress]{revtex4}
\usepackage{amssymb,amsmath}
\usepackage{graphicx}

\def\R{{\mathbb{R}}}      
\def\C{{\mathbb{C}}}      
\def\H{\mathcal{H}}       
\def\RR{\mathcal{R}}
\def\L{\mathcal{L}}

\def\O{\mathcal{O}}

\def\a{{\hat a}}

\def\x{{\hat x}}
\def\N{\hat N}
\def\q{\hat q}
\def\p{\hat p}
\def\f{\hat f}
\def\g{\hat g}
\def\cf{{\it cf. }}
\def\eg{{e.g.\ }}
\def\wig{_{\rm\scriptstyle w}}
\def\NN{{\mathbb{N}}}      
\def\Re{\mbox {\rm Re} }
\def\Im{\mbox {\rm Im} }
\begin{document}

\title{Moyal phase-space analysis of nonlinear optical Kerr media}

\author{T. A. Osborn}
\affiliation{Department of Physics and Astronomy,
University of Manitoba,
Winnipeg, Manitoba R3T 2N2, Canada}

\author{Karl-Peter Marzlin}
\affiliation{Department of Physics, St. Francis Xavier University,
  Antigonish, Nova Scotia, B2G 2W5, Canada}
\affiliation{Department of Physics and Astronomy,
        University of Calgary, Calgary, Alberta T2N 1N4, Canada}

\begin{abstract}
Nonlinear optical media of Kerr type are described by a particular version of an anharmonic quantum harmonic oscillator. The dynamics of this system can be described using the Moyal equations of motion, which correspond to a quantum phase space representation of the Heisenberg equations of motion. For the Kerr system  we derive exact solutions of the Moyal equations for a complete set of observables formed from the photon creation and annihilation operators. These Moyal solutions incorporate the asymptotics of the classical limit in a simple explicit form. An unusual feature of these solutions is that they exhibit periodic singularities in the time variable. These singularities are removed by the phase space averaging required to construct the expectation value for an arbitrary initial state. Nevertheless, for strongly number-squeezed initial states the effects of the singularity remain observable.
\end{abstract}

\maketitle


\section{Introduction}

Phase space methods provide one of the most important tools to investigate the
relation between classical and quantum mechanics. In classical dynamics, the
state of a system can be described by a probability distribution that is a
function of $x \equiv (q,p)$, i.e., of position $q$ and momentum $p$ of a
particle. In quantum physics such a distribution cannot exist because
Heisenberg's uncertainty relation prohibits simultaneous knowledge of $q$ and
$p$, but a number of quasi-distributions have been proposed for a phase space
analysis of quantum systems \cite{wallsMilburn}. Among the most popular is the
Wigner function which is related to the density matrix $\rho(q,q') = \langle q
|\hat{\rho} |q' \rangle $ of a particle by
\begin{equation}
  W(x) =  \frac{1}{\pi\xi} \int_{-\infty}^\infty  \; \rho(q-q',
  q+q')\, e^{\,2i p q'/\xi} \,dq'\; .
\label{eq:wignerfunction} \end{equation}
Usually the parameter $\xi$ is
replaced by $\hbar$, but for reasons explained below we use $\xi$ instead. The
Wigner function is not a probability distribution because it can have negative
values, which are often interpreted as an indication for genuine quantum
effects.

The Wigner function and its time evolution for a given Hamiltonian have been applied to analyze a huge variety of phenomena; in particular during the last two decades it has been used to analyze the quantum state of light \cite{lvovskyReview}. However, to investigate the classical-quantum correspondence the closely related concept of the Weyl symbol $[\widehat{A}]\wig$ of an operator $\hat{A}$ may be more suitable. The Weyl symbol has essentially the same definition as the Wigner function (\ref{eq:wignerfunction}), with $\widehat{\rho}$ replaced by a general operator $\widehat{A}$. For the special case of the density matrix it leads to the relation $[\hat{\rho}]\wig = 2 \pi \xi\, W(x)$. The difference between the Weyl symbol and the Wigner function is that the latter refers to the quantum state while the former refers to observables. For an observable, the time evolution of its Weyl symbol is governed by the Moyal equation \begin{equation}\label{mo1} \frac{d}{dt} A\wig(t) = \left\{A\wig(t) , H\wig \right\}_M\,. \end{equation} Above $H\wig, A\wig(t)$ are the Weyl symbols of the Hamiltonian and $\widehat{A}(t)$, respectively. The Moyal bracket $\{\cdot,\cdot \}_M$, \cf (\ref{Moyal1}),  is the quantum extension of the classical Poisson bracket. The Moyal equation corresponds to a phase space formulation of the operator valued Heisenberg equations of motion. The similarity between the Heisenberg equation for observables and the classical equations of motion makes the Weyl symbol representation a powerful tool to shed light on the relation between classical and quantum dynamics. The Moyal equation of motion enables one to study the dynamics of observables without reference to the quantum state of the system.

Despite these advantages the Moyal equation has not been used extensively to analyze quantum systems because it is considerably more difficult to solve than the Schr\"odinger equation. For this reason not many exact solutions are known, and a comparison with experimental data is often difficult. The purpose of this paper is to improve this situation by providing an exact solution of the Moyal equation for an experimentally relevant system: the Kerr model of nonlinear optics for a single-mode of the quantized radiation field. We obtain analytical solutions of the Moyal equations that are valid for all time. These exact solutions are then used to characterize the transition from classical to quantum dynamics and to provide a phase-space based physical interpretation of the Kerr effect.

This paper is organized as follows. In Sec.~\ref{sec:kerr} we review the basic properties of the Kerr effect. A closed, exact expression (\ref{MA1}) for the phase space representation of all Heisenberg operators in the quantized Kerr model is derived in Sec.~\ref{sec:moyalkerr} and its classical limit is studied in Sec.~\ref{sec:class}. In Sec.~\ref{sec-expval} we show that the periodic divergences in the Moyal representation of the annihilation operator generate characteristic, finite peaks in the expectation values of the canonical variables for squeezed coherent states of light. In Sec.~\ref{sec-expval2} it is shown that the expectation values of all observables remain finite for general initial states.

\section{The Kerr model}\label{sec:kerr}
The Kerr model of optical nonlinearities is one of the most studied systems in quantum optics. In a Kerr medium, the refractive index of a classical beam of light depends on the light intensity as $n = n^{(0)} + n^{(2)} I$, where $n$ denotes the total refractive index, $n^{(0)}$ the linear refractive index, $I$ the light intensity, and~$n^{(2)}$ the optical Kerr coefficient. The Kerr effect is invaluable for spectral broadening  and self-focusing of laser pulses~\cite{shen}. It usually appears in special crystals and its magnitude is typically so small that large light intensities are needed. However, recent research on electromagnetically induced transparency \cite{Harris97,PRL66:2593,Lukin97} has made very large Kerr coefficients with values of up to $ n^{(2)}\approx 0.1 \text{cm}^2$/W  possible \cite{Schmidt96,Harris99,Hau99,Bajcsy03} and may even lead to nonlinear effects at the single-photon level \cite{Lukin00,Pet02,Matsko03,Pet04,Wang06}.

A quantum description of the Kerr effect can be accomplished by replacing the intensity of light in the refractive index by the corresponding operator. This is equivalent to introducing a quartic interaction term in the radiation Hamiltonian \cite{wallsMilburn}. A particularly simple description can be achieved if the photon dynamics is confined by an optical cavity with high finesse mirrors. Such cavities may support only a single light mode in a given spectral range, so that the dynamics can be described by operators $\hat{a}, \hat{a}^\dagger $ that annihilate or create a photon in the cavity mode, respectively. The photon number operator is given by $\widehat N = {\hat a}^\dag \hat a$, and the Kerr Hamiltonian is given by the Wick ordered operator
\begin{equation}\label{2.H1}
\widehat H = \omega_2 ({\hat a}^\dag)^2 {\hat a}^2 + \omega_1 {\hat
a}^\dag \hat{a} \; .
\end{equation}
Physically, $\omega_1$ is related to the linear index of refraction by $\omega_1 = kv_\text{gr}  n^{(0)} $ and $\omega_2$ to the nonlinear refractive index by $\omega_2 = k  v_\text{gr}n^{(2)}I_0$, where $v_\text{gr}$ is the group velocity of light in the medium, $k$ its wave number, and $I_0 = 2\hbar k c^2/V$ the intensity of a single photon in a cavity of volume $V$. The operators $\widehat H, \hat{a}$ act on the Hilbert space $\H=L^2(\R,\C)$ and satisfy harmonic oscillator commutation relations
\begin{equation}\label{alg1}
   [\a , \hat{a}^\dagger ] = \xi I \qquad [\N,\a] = -\xi\a \qquad [\N,\hat{a}^\dagger ] =
   \xi \hat{a}^\dagger \,.
\end{equation}
With this notation we have introduced the real dimensionless parameter $\xi$ that allows us to interpolate between classical and quantum evolution. The fundamental distinction between quantum and classical mechanics resides in commutivity. The product of observables in classical mechanics is abelian whereas the product operation in quantum mechanics is noncommutative. As $\xi \rightarrow 0$, noncommuting behavior in the Kerr model is suppressed, and for $\xi=1$ standard single mode photon physics is recovered. Specifically, the Moyal equation of motion automatically incorporates Bohr's correspondence principle: in the limit $\xi\rightarrow 0$, the Moyal bracket becomes the Poisson bracket and Eq.~(\ref{mo1}) then turns into the Poisson equation of motion.

 The relation between $\xi$ and the conventional ``quantization parameter'' $\hbar$ can be seen by relating the creation and annihilation operator to two Hermitian operators via  $\hat{a} = (\hat{q}+i\hat{p})/{\sqrt 2}$ and $\hat{a}^\dagger = (\hat{q}-i\hat{p})/{\sqrt2}$. For dimensionless position and momentum operators $\hat x \equiv(\q,\p)=(\hat{x}_1,\hat{x}_2)$, one then has
\begin{equation}\label{heis1}
    [\hat{x}_j,\hat{x}_k] = i\xi J_{jk} \qquad J =\left(
                                          \begin{array}{cc}
                                            0 & 1 \\
                                            -1 & 0 \\
                                          \end{array}
                                        \right)\; ,
\end{equation}
where $J$ is the Poisson matrix.
In a coordinate parametrization where $\q$ is proportional to length and $\p$ to momentum, $[\hat{x}_j,\hat{x}_k] = i\hbar J_{jk}$, so the limit $\xi \rightarrow 0$ is equivalent to letting Planck's constant $\hbar$ go to zero for a conventional quantum harmonic oscillator.  We use $\xi$ instead of $\hbar$ because in quantum optics the operators  $\hat{a} , \hat{a}^\dagger$ have a different physical interpretation than for a single Schr\"odinger particle in a harmonic potential. As a consequence, their degree of commutativity is not controlled by $\hbar$, but rather by the mathematically introduced deformation parameter $\xi$.

Throughout our derivations free use is made of the Weyl symbol calculus that represents Hilbert space operators by functions in phase space.  An overview of this quantum phase space representation and its non-commutative $\star$ product is presented in the Appendix A.


\section{The Moyal-Kerr Problem and its Solution}\label{sec:moyalkerr}

In this section we describe Heisenberg picture evolution in Weyl symbol form, identify the symmetries of the Moyal equation of motion and use these symmetries to construct an exact solution.\smallskip

The Weyl symbol representation
of the Hamiltonian (\ref{2.H1}) is
\begin{equation}\label{2.H2}
        [\widehat H]_{{\rm\scriptstyle w}} \equiv H(\xi,x) =
        \omega_2 \left[\frac 14 x^4  - \xi x^2 + \frac 12 \xi^2
        \right] + \omega_1\left[ \frac 12 x^2 - \frac 12 \xi \right]\,,
\end{equation} where $x^2=q^2+p^2$.
The $\omega_2$ term is the phase space form of the non-linear interaction. The $\omega_1$ portion is the symbol of the number operator $[\N]_{{\rm\scriptstyle w}}(x)=N(x)=\frac 12 (x^2 - \xi)$ and represents the evolution of non-interacting photons.

First consider the Moyal equation for a general observable. Denote
Schr\"{o}dinger evolution by $U_t=\exp (-i t\widehat H/\xi)$. Let
$\widehat{\Theta}_0$ be an observable with dynamical value
$\widehat{\Theta}(t) = U_t^\dag\, \widehat{\Theta}_0\, U_t$ and
Heisenberg equation
\begin{equation}\label{2.Heis}
    \frac d{dt} \widehat{\Theta}(t) = i\xi^{-1}
    [\widehat{H},\widehat{\Theta}(t)]\,.
\end{equation}
The Weyl symbol image of Eq.~(\ref{2.Heis}) is Moyal's equation (\ref{mo1}). Let $\Theta(t|x)\equiv[\widehat{\Theta}(t)]_{{\rm\scriptstyle w}} (x)$ be the symbol of the evolving observable, then
\begin{eqnarray}
\label{mo2} \nonumber \dot \Theta(t|x) &=&  \{\Theta(t), H\}_M(x) =
i\xi^{-1}\left( H\star \Theta(t) - \Theta(t)
    \star H\right) (x) \\ \label{mo3}
   &=&  i\xi^{-1}\left(H(\L) - H(\RR)\right)\, \Theta(t|x)\,.
\end{eqnarray}
In the last identity, we employ the expression of the Moyal bracket in terms of the left and right operators $\L$ and $\RR$ which are defined in Eq.~(\ref{LR1}). It converts the Moyal bracket into a differential operator acting on the target function $\Theta(t|x)$. Evaluating, $H(\L) - H(\RR)$ for Hamiltonian~(\ref{2.H2}) one obtains the following third order differential equation
\begin{equation}\label{2.M2}
  \dot \Theta(t|x) = -\left[\omega_2\bigg(x^2 -2\xi- \frac {\xi^2}{4}
  \partial_x^2\bigg) +\omega_1 \right]
  (x \cdot J \partial_x) \Theta(t|x)\,.
\end{equation}

To fully characterize a quantum system, Eq.~(\ref{2.M2}) has to be solved for a complete set of operators. For the Kerr-Moyal problem, such a set is given by $\{(\hat{a}^\dagger )^s \a^m | 0\leq s,m \in \NN \}$. We denote by
\begin{equation}
  \Theta_{sm}(t|x)\equiv[(\hat{a}(t)^\dagger )^s \a(t)^m]_{{\rm\scriptstyle w}}(x)
\end{equation}
the Weyl symbol of the corresponding operators in Heisenberg picture, where $\hat a(t)= U_t^\dag\, \hat a\, U_t$. Our task is to solve Eq.~(\ref{2.M2}) for the set of symbols $\Theta_{sm}(t|x)$ with initial conditions $\Theta_{sm}(0|x)\equiv[(\hat{a}^\dagger(0) )^s \a^m(0)]_{{\rm\scriptstyle w}}(x)$.

Moyal equation (\ref{2.M2}) has the form of a Schr\"{o}dinger equation over the $x$-variable manifold. Specifically, the function $\Theta_{sm}(t|x)$ may be considered an unnormalized `state' over the manifold $T^*\R=\R^2$. This is a general feature of the Moyal equation and has been used to construct {\small{WKB}} type asymptotic approximations \cite{OK02} for $\Theta(t|x)$. The system (\ref{2.M2}) admits a standard \cite{MF81} small $\xi$ expansion because the highest order differential operator, the Laplacian $\partial_x^2$, is scaled by $\xi^2$.

Quantities like $\L$ and $\RR$ act on the Weyl symbols $\Theta_{sm}(t|x)$, while operators like $\hat{a}$ act on the usual Hilbert space. To distinguish between these two cases we use a hat to denote the latter and script capital letters denote operators acting on the Hilbert space $\H_2=L^2(\R^2,d^2x)$.

\smallskip
It is now useful to determine the symmetries present in the equation of motion
(\ref{2.M2}). In this context it is advantageous to introduce complex
coordinates $z=(q+ip)$ and $\partial_z = \frac{1}{2} (\partial_q -
i\partial_p)$. In this notation $[\a]\wig(x)= a(x) = z/{\sqrt 2}$ and
$[\a^\dag]\wig(x)= a(x)^* = z^*/{\sqrt 2}$.   Because $a$ is a linear
function: $a\star a = a^2$, etc., giving
\begin{equation*}
  [\a^m]_{{\rm\scriptstyle w}}(x) = 2^{-m/2} z^m \,, \quad\quad
  [(\a^\dagger )^s]_{{\rm\scriptstyle w}}(x) = 2^{-s/2}
  z^{*s}\,.
  \label{a0W}\end{equation*}
 Employing Eq.~(\ref{LR2})
it is then straightforward to show that the initial condition is
\begin{equation}\begin{split}
  \Theta_{sm}(0|x) &=
   \overline{a}(\L)^s a^m(x) =
  \left[\frac{1}{\sqrt{2}}(z^* - \xi
  \partial_z)\right]^s \left({\frac z{\sqrt 2}}\right)^m
\label{initCond1} \\
   &= \sum_{l=0}^{\min(s,m)} W(m,s,l) \big(-\frac{\xi}{2}\big)^l\,
   \overline{a}^{s-l}(x)\, a^{m-l}(x)
 \end{split}\end{equation} with $W(m,s,l) \equiv
s!m!\left[l!(s-l)!(m-l)!\right]^{-1}$. For the set of operators under
consideration, the Kerr-Moyal equation (\ref{2.M2}) then becomes
\begin{equation}
  \dot \Theta_{sm} (t|x) = -\left[\omega_2 {\cal K} +\omega_1 \right]
  (x \cdot J \partial_x) \Theta_{sm} (t|x)
\label{2a.M2} \end{equation}
with
\begin{eqnarray*} &{\cal K}& \equiv  |z|^2 -2\xi- \xi^2 \partial_z
\partial_{z^*} = x^2 -2\xi- \xi^2 \partial_x^2 \vphantom{\Bigg |} \\ &(x& \cdot
J \partial_x) = i (z \partial_z - z^* \partial_{z^*})\; . \end{eqnarray*}
The action of the phase space operator $(x \cdot J \partial_x)$ is similar to that of the angular momentum operator $\hat{L}_z$ on Hilbert space. This can be easily seen in polar coordinates, $z = r e^{i\phi}$, where it takes the form $(x\cdot J \partial_x) = \partial_\phi$ . Furthermore, the initial condition $\Theta_{sm}(0|x)$ is an eigenstate of  $(x \cdot J \partial_x)$ with eigenvalue $\lambda_{sm} = i(m-s)$. Because $[ {\cal K}, (x \cdot J \partial_x)] =0$ we can infer that $\Theta_{sm}(t|x)$ will remain an eigenstate of $(x \cdot J \partial_x)$ with the same eigenvalue. Via this eigenfunction mechanism the third order partial differential equation (\ref{2a.M2}) is reduced to second order,

\begin{equation}\label{EQM}
    \dot \Theta_{sm} (t|x) = -i(m-s)\left[\omega_2 {\cal K} +\omega_1 \right]
   \Theta_{sm} (t|x)\,.
\end{equation}

We remark that if $m=s$ the right side of Eq.~(\ref{EQM}) is zero.
This means that $\Theta_{mm}(t|x) = (x^2/2)^m$ is a constant of motion. This
function is also a classical constant of motion because $ \{ H, x^{(2m)}\} = 0$.

Equation ~(\ref{EQM}) has the formal solution
\begin{equation}
   \Theta_{sm}(t|x) =e^{-i(m-s)\omega_1 t}
   e^{ -i(m-s)t \omega_2 {\cal K}}  \Theta_{sm}(0|x) \; .
\end{equation}
We can now take advantage of the special form (\ref{initCond1}) of the initial
conditions. It is well known \cf (\cite{merz}, pg. 40) that $e^{\hat{K}}
\hat{A} = e^r \hat{A} e^{\hat{K}}$ for $[\hat{K},\hat{A}] = r\hat{A}$ and $r
\in \C$. Because
\begin{equation*}
  \left [ {\cal K} ,  \frac{1}{\sqrt{2}}(z^* - \xi
  \partial_z) \right ] =
  \xi \frac{1}{\sqrt{2}}(z^* - \xi \partial_z)
\end{equation*}
we can express the formal solution as
\begin{eqnarray*}
  \Theta_{sm}(t|x) &=&
 2^{-s/2}
  \exp \big ( -i(m-s) t ( \omega_1 + \omega_2 \xi s) \big ) \,
  (z^* - \xi \partial_z)^s
  e^{ -i(m-s)t \omega_2 {\cal K}} \left ( \frac{z}{\sqrt{2}} \right
  )^m
\\ &=&
   2^{-s/2}
  \exp \big ( -i(m-s) t \omega_2 \xi s \big ) \,
  (z^* - \xi \partial_z)^s\,
  \Theta_{0m}(m^{-1}(m-s)t|x) \; .
\label{formalSol1}\end{eqnarray*} It is therefore sufficient to find a closed
form for $ \Theta_{0m}(t|x) $. To do so we make the ansatz
\begin{eqnarray}
\Theta_{0m}(t|x) &=& e^{g(t) x^2} e^{-im (\omega_1 - 2
  \xi\omega_2) t} f(t) \,  a^{m}(x) \; ,
 \label{A1a}
\end{eqnarray}
with initial conditions $g(0)=0$ and $f(0)=1$.
Inserting this ansatz into Eq.~(\ref{EQM}) and sorting the resulting
equation in powers of $|z|^2$ yields a coupled set of differential
equations for $g(t)$ and $f(t)$,
\begin{eqnarray*}
\dot g = i m \omega_2 (-1+\xi^2 g^2) \qquad
\dot f  = i m (m+1) \xi^2 \omega_2 g  f  \; ,
\end{eqnarray*}
which have the solutions
\begin{eqnarray*}
  g(t) = -\frac{i}{\xi} \tan (m\xi \omega_2 t) \qquad
  f(t) =   \big(\sec(m\xi \omega_2 t)\big)^{m+1} \,.
\end{eqnarray*}

This leads to one of the main results of this work: the exact solution of the
Moyal equation for the complete set of operators $(\hat{a}^\dagger)^s
\hat{a}^m$ is given by
\begin{eqnarray}
  \Theta_{sm}(t|x) &=&
   e^{-i(m-s) \omega_1 t} \,  e^{ i (2-s) \tilde{t}}
  ( \sec \tilde{t}\,)^{m+1} \,
  \left ( \frac{ z^* - \xi \partial_z}{\sqrt{2}} \right )^s
  \exp \left (
  -\frac{i}{\xi} z z^* \tan \tilde{t}
  \right )
   \left ( \frac{ z}{\sqrt{2}}\right )^m
\label{MA1} \end{eqnarray} with $\tilde{t} \equiv (m-s)\xi \omega_2 t$.
Evaluating the $s$-fold derivative and converting to the phase space variables
$x$ yields
\begin{equation}\label{MA2}
    \Theta_{sm}(t|x) =
   e^{-i(m-s) \omega_1 t}
  (\sec \tilde{t}\,)^{s+m+1} \,
    \exp \Big( 2 i \, \tilde{t}
  -i\frac{x^2}{\xi}  \tan \tilde{t}
  \Big )
   \sum_{l=0}^{min(s,m)} W(m,s,l) \Big(-\frac{\xi}{2} e^{-i\tilde{t}}\cos \tilde{t}\, \Big)^l\,
   \overline{a}^{s-l}(x)\, a^{m-l}(x)\,.
\end{equation}
 This final form displays the adjoint symmetry:
 $[(\hat{a}(t)^\dagger )^s \a(t)^m]^\dag = (\hat{a}(t)^\dagger )^m \a(t)^s$,
  or equivalently ${\Theta_{sm}(t|x)}^* = \Theta_{ms}(t|x)$. \vskip 1mm

A striking feature of the solutions  (\ref{MA2}) is that they have  a singular amplitude for times whenever $\cos \tilde{t} =0$. Henceforth we will refer to this behavior as the Moyal singularity. Its mathematical origin is that $g(t)$ obeys a non-linear Ricatti equation.

\section{Classical and Quantum Trajectories}\label{sec:class}

The manner in which quantum phase space solutions embed the classical dynamics occurs in two different ways.  In the first way, one characterizes how the solutions $\Theta_{sm}(t|x)$ transform into the Poisson equation solutions as $\xi \rightarrow 0$.  The second semiclassical association relates quantum expectation values to corresponding classical flows. The first way, the phase space correspondence, is treated in this section.

Quantum trajectories on phase space are defined as the symbol image of the Heisenberg coordinate operator evolution:
$\hat{x}(t)=U_t^\dag \hat{x}\, U_t$, in detail
\begin{equation}\label{3A}
Z(t,\xi|x) \equiv [\hat{x}(t)]_{{\rm\scriptstyle w}}(x) = \left( [\q(t)]_{{\rm\scriptstyle w}}, [\p(t)]_{{\rm\scriptstyle w}}]\right)(x)\,.
\end{equation}
The Moyal solutions above give formulas for  $Z(t,\xi|x)$ via the relationships
\begin{eqnarray*}
  [\q(t)]_{{\rm\scriptstyle w}}(x) &=& \frac{1}{\sqrt{2}} \big(\Theta_{01}(t|x) + \Theta_{10}(t|x)\big)
  = \sqrt 2\, \Re\, \Theta_{01}(t|x) \label{3B1} \\
  {\phantom a} [\p(t)]_{{\rm\scriptstyle w}}(x)&=& \frac{1}{\sqrt{2}} \big(\Theta_{01}(t|x) -
  \Theta_{10}(t|x)\big) = \sqrt 2\, \Im\,
  \Theta_{01}(t|x)\,. \label{3B2}
\end{eqnarray*}

The dynamics for a the classical version of the Kerr problem is simple. The
$\xi =0$ part of $H(\xi,x)$ defines the classical Hamiltonian
\begin{eqnarray*}
 \label{3C1}
  &H(\xi,x) = H_\text{cl}(x) + \xi h_1(x) + \frac {\xi^2}{2!}h_2(x) \\ \phantom{\Big |} \label{3C2}
  &H_\text{cl}(x)  = \frac 14 \omega_2 x^4 + \frac 12 \omega_1 x^2 \qquad
  h_1(x)= -\omega_2 x^2 \qquad h_2(x) = \frac 12 \omega_2\,.
\end{eqnarray*}
The classical trajectory $Z_\text{cl}(t|x) = \big(q_\text{cl}(t|x),
p_\text{cl}(t|x)\big)$ is then the solution of  Hamilton's equation
\begin{equation*}\label{3D}
    \dot Z_\text{cl}(t|x) = J \partial_x
    H_\text{cl}\big(Z_\text{cl}(t|x)\big)
\end{equation*} with initial condition $Z_\text{cl}(0|x)=x$.
The solution is $Z_\text{cl}(t|x) = [\,\exp t(\omega_2 x^2+
\omega_1) J\,]\,x$. In matrix form this is
\begin{equation}\label{3E}
    Z_\text{cl}(t|x) = \left(
               \begin{array}{c}
                 q_\text{cl}(t|x) \\
                 p_\text{cl}(t|x) \\
               \end{array}
             \right) =
    \left(
      \begin{array}{cc}
        \cos(\omega_2 x^2 + \omega_1 )\,t & \sin(\omega_2 x^2 + \omega_1 )\,t \\
        -\sin(\omega_2 x^2 + \omega_1 )\,t & \cos(\omega_2 x^2 + \omega_1 )\,t \\
      \end{array}
    \right)
    \left(
      \begin{array}{c}
        q \\
        p \\
      \end{array}
    \right).
\end{equation}
This is oscillatory motion with a variable  frequency $\omega_2 x^2 + \omega_1$
that depends on the  initial value, $x$. The frequency increases as the
constant of motion $x^2$ increases.

To compare the classical and the quantum trajectory it is useful to
introduce the complex quantity
\begin{equation}\label{3Fa}
  a_\text{cl}(t|x) \equiv \frac{ 1}{\sqrt{2}} \Big ( q_\text{cl}(t|x) + i
  p_\text{cl}(t|x) \Big )
  = e^{-i(\omega_2 x^2 + \omega_1)\,t }\,\frac{ 1}{\sqrt{2}} (q+i p)
  \,,
\end{equation}
which is the classical quantity corresponding to the annihilation operator. It agrees with the predictions of the Kerr model for the complex electric field amplitude of classical light: the leading phase factor represents the phase shift that light would experience when it travels through a nonlinear medium of length $L= t v_\text{gr}$. In an optical system, $x^2$ represents the mean number of photons in the cavity, which can also be expressed as the intensity of the light field in units of the intensity of a single photon in the cavity.

Employing Eq.~(\ref{3E}) we can express the related
quantum trajectory as
 \begin{equation}\label{3G}
    \Theta_{01}(t|x) = \sec^2(\xi \omega_2 t)\, e^{i  \Phi(\xi,x,t)}
    a_\text{cl}(t|x) \; ,
 \end{equation}
with the quantum phase factor
\begin{equation}\label{3F}
        \Phi(\xi,x,t) \equiv 2\xi \omega_2 t + x^2\big(\omega_2 t -
        \xi^{-1} \tan (\xi \omega_2 t)\big) \,.
    \end{equation}
This phase vanishes at $\xi= 0$.
If one implements a power series expansion of (\ref{3G}) about $\xi=0$, the result defines the semiclassical expansion of the Moyal solution.  To first order in $\xi$ one has
\begin{equation}\label{SC1}
    \Theta_{01}^{\text{sc}}(t|x) = a_\text{cl}(t|x)\big[ 1 +2i\omega_2t\,\xi +O(\xi^2) \big]\,.
\end{equation}

Expressions (\ref{3G}-\ref{SC1}) show how the periodic quantum and classical flows are interdependent.
If $|\xi\omega_2 t| \ll \pi/2$ and $|\Phi(\xi,x,t)| \ll \pi/2$ the quantum and classical trajectories nearly coincide. This limit very well describes all experiments with conventional nonlinear optical crystals for which the nonlinear refractive index $n^{(2)}$ is very small.

At the other extreme, when $\xi\omega_2 t$ is close to an odd multiple of $\pi/2$,
the amplitude factor $\sec^2(\xi\omega_2 t)$
is diverging and the quantum phase rotation $\Phi(\xi,x,t)$ is undergoing near infinite oscillation. This regime should soon
be experimentally accessible by using EIT-based nonlinear media \cite{Schmidt96,Harris99,Hau99,Bajcsy03}, but below we will
explore if such large oscillations are actually observable.

\begin{figure}
\includegraphics[width=8cm]{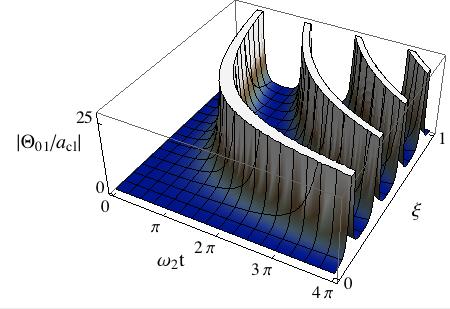}
\caption{\label{fig-qampl}
  Magnitude of the ratio between the quantum amplitude $\Theta_{01}$
  and the classical amplitude $a_\text{cl}$.}
\end{figure}

\begin{figure}
\includegraphics[width=8cm]{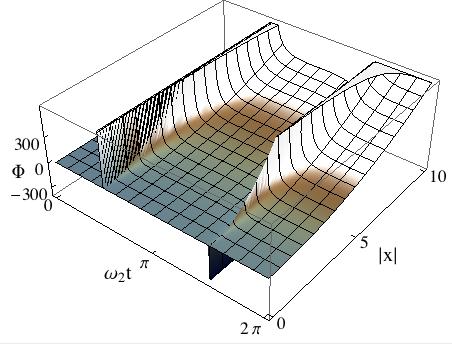}
\caption{\label{fig-qphase}
 Quantum phase factor $\Phi (\xi, x,t)$ for $\xi=1$. }
\end{figure}

Fig.~\ref{fig-qampl} shows the quantum effect of the nonlinearity on the complex field amplitude $|a_\text{cl}(t|x)|$, which is independent of $x^2$. It can clearly be seen that the periodic divergences disappear in the classical limit $\xi \rightarrow 0$, and that even for a fully quantized theory ($\xi=1$) they only appear for very large nonlinear refractive indices.
Fig.~\ref{fig-qphase} displays the quantum phase factor $\Phi$ for a fully
quantized theory ($\xi=1$). Like the amplitude it displays a periodic
divergence in time. The width of the divergences in phase space is proportional
to $x^2$, indicating that they are an intensity-dependent effect. The
divergences disappear in the classical limit $\xi\rightarrow 0$.

Generally, not everything is rapidly oscillating. Recall that $x^2$ is both a classical and quantum constant of motion.  At the classical level this constant is recovered from the flow via $Z_{cl}(t,\xi|x)^2 = x^2$. At the quantum level one has $Z(t,\xi)\star Z(t,\xi)(x)= x^2$. This latter identity can be derived from Berezin's representation (\ref{Ber1}) of the star product which leads to a four dimensional Fresnel integral with value $x^2$.

In the literature \cite{OM95,McQ98} the quantum trajectory $Z(t,\xi|x)$ is
generally approximated by a small $\xi$ asymptotic approximation
\begin{equation}\label{3J}
Z(t,\xi|x) = Z_\text{cl}(t|x) + \xi {\it z}^{(1)}(t|x) + \frac {\xi^2}{2!} {\it z}^{(2)}(t|x) +
\cdots
\end{equation}
In the Kerr--Moyal problem, and for most problems, this expansion has terms to all order in $\xi$. The exception occurs when $H$ is quadratic. In this special case just the leading term $Z_\text{cl}(t|x)$ is nonzero. For this reason little can be learned about the general nature of the classical-quantum transition by investigating quadratic Hamiltonian problems.  This feature is seen in Moyal solutions above. The effect of the  $x$-quadratic part of Hamiltonian (\ref{2.H2}) on $\Theta_{sm}(t|x)$ is confined to the phase factor $\exp(-i(m-s)\omega_1 t)$. This factor has no $\xi$ or $x$ dependence.

If an exact formula for $Z(t,\xi|x)$ is known then these higher order expansion
coefficients are given by
\begin{equation*}\label{3K}
{\it z}^{(n)}(t|x) = \frac {\partial^n}{{\partial \xi}^n} Z(t,\xi|x)\Big
|_{\xi=0}\,.
\end{equation*}
For example, the leading semiclassical correction to the classical Kerr problem
flow is
\begin{equation}\label{3L}
{\it z}^{(1)}(t|x) = \frac {\partial}{{\partial \xi}} Z(t,\xi|x)\Big
|_{\xi=0} = - 2\, \omega_2 t\, Z_\text{cl}(t|x)\,.
\end{equation}

The method for computing $z^{(n)}(t|x)$ when the full quantum trajectory is not available is to expand the Moyal equation identity in powers of $\xi$. This approach works if both the observable and the Weyl system Hamiltonian are semiclassical admissible, namely both admit a power series expansion  about $\xi=0$. This is the situation in the Kerr problem. The $\O(\xi^1)$ portion of the Moyal equation for $Z(t,\xi|x)$ is an inhomogeneous Jacobi field equation for the unknown ${\it z}^{(1)}(t|x)$, i.~e.
 \begin{equation}\label{3M} \left[ {\frac d{dt}} - J H_\text{cl}^{''}\big( Z_\text{cl}(t|x) \big)\right] {\it z}^{(1)}(t|x) = J \partial_x  h_{1}(Z_\text{cl}(t|x)) \end{equation}
  with initial condition ${\it z}^{(1)}(0|x)=0$. The quantity $H_{c}^{''}$ is the Hessian matrix of $H_c$.  Similar equations define the higher order corrections ${\it z}^{(n)}(t|x)$. A Jacobi field is a solution the homogenous version of Eq.~(\ref{3M}) and provides a linearized prediction for small deviations about the classical flow $Z_\text{cl}(t|x)$.

One can readily check that $z^{(1)}(t|x)= - 2\, \omega_2 t\, Z_\text{cl}(t|x)$ is a solution of Eq.~(\ref{3M}). This demonstrates the compatibility of $Z(t,\xi|x)$ in Eq.~(\ref{3A}) with the standard asymptotic semiclassical expansion generated by Eq.~(\ref{3J}).  Formula (\ref{3L}) also illustrates the small time limitation of this expansion.  The ${\it z}^{(1)}$ correction has unbounded growth in $t$; in order for the correction to be small one requires $|2\, \omega_2 t| \ll 1$. The next correction term $z^{(2)}$ grows like $x^2(\omega_2 t)^3$. This shows that the expansion (\ref{3J}) is non-uniform in the $(t,x) \in \R\times\R^2$ domain.

\section{Dynamical Expectation Values and their Classical Limit}\label{sec-expval}

In this section we compute the squeezed state expectation value of the Moyal solution corresponding to $\q(t)$ and $\p(t)$ and characterize their semiclassical limits. The squeezed states are of particular interest because they will allow us to study the effects of the singularity. Furthermore, squeezed states are of high practical value because they correspond to non-classical states of light which can be used for quantum information \cite{appel:093602}. In addition, the reduced noise of specific observables makes them of interest in high-precision experiments such as gravitational wave interferometers \cite{McKenzie2006}
\smallskip

Squeezed states are unitary modifications of coherent states. We recall the
defining equations for coherent and squeezed states. The coherent states
$|\alpha \rangle$, are translated vacuum states (see, e.g.,
Ref.~\cite{wallsMilburn}). The translation operator $D(\alpha),\, \alpha \in\C,\ \arg \alpha \in [0,2\pi)$
shifts $\a$ by
\begin{eqnarray*}
  D^\dag (\alpha) \a D(\alpha) = \a + \alpha I
\quad , \quad
 D(\alpha) \equiv
  \exp\big[\xi^{-1} (\alpha \hat{a}^\dagger  - \alpha^* \a) \big]\,.
\end{eqnarray*}
Defining  $|\alpha\rangle = D(\alpha) |0\rangle$,  it follows that $\a
|\alpha\rangle = \alpha |\alpha\rangle$. The states $|\alpha \rangle$ have unit
normalization with inner product  $\langle \alpha| \beta
\rangle = \exp\big[ -\xi^{-1}(\frac 12 |\alpha|^2 + \frac 12 |\beta|^2 -
\alpha^*\beta)\big] \phantom{ \Big |}$.

Given the coherent states one obtains the squeezed states by the action of a
unitary Bogoliubov operator,
$V(\tau) \equiv \exp [\tau(\hat{a}^\dagger )^2 -\tau^* \a^2], \ \tau =|\tau|\exp(i\phi), \ \phi \in [0,2\pi)$.
The squeezed states $|\tau\alpha \rangle \equiv V(\tau) | \alpha \rangle$ are the eigenfunctions of
\begin{equation*}\label{4C}
V(\tau)\, \a V(\tau)^\dag |\tau\alpha \rangle = \alpha |\tau\alpha \rangle\,.
\end{equation*} Because $V(\tau)$ is unitary, the coherent state eigenvalue $\alpha$ and normalization
$\langle \tau\alpha | \tau \beta \rangle = \langle \alpha | \beta \rangle $ are unchanged.

The coherent and squeezed states are interpreted as being near classical
because they have special properties with respect to the uncertainty relations.
The squeezed state mean values are readily found by exploiting the metaplectic
nature of the Bogoliubov transform, specifically,
\begin{equation}\label{B1}
    V(\tau)  \x V(\tau)^\dag = S(\tau)\, \x\qquad\quad S(\tau)=\left(
              \begin{array}{cc}
                 s \cos^2(\phi/2) + s^{-1} \sin^2(\phi/2)  & -\frac 12\, ( s^{-1}-s)\sin(\phi)\\
                -\frac 12\, ( s^{-1}-s)\sin(\phi) & \quad s^{-1} \cos^2(\phi/2) + s \sin^2(\phi/2)
               \vphantom{ \Big ]} \\
              \end{array}
            \right) .
\end{equation}
The parameter $s \equiv \exp (-2 \xi |\tau| ) \leq 1 $ describes the amount by
which the uncertainty of a canonical variable can be reduced (see
Eq.~(\ref{V1})). The set $\{S(\tau)|\tau\in \C\}$ is a family of positive
symplectic matrices with inverse$S(\tau)^{-1} =
S(-\tau)$ and group multiplication law $S(\tau)^{\,2} = S(2\tau)$.

The $q,p$ variances  turn out to be \cite{Stoler70}
\begin{equation}\label{V1}
  \langle \Delta q\rangle_{\tau\alpha}^2 = \frac \xi 2\, \left ( \frac{ 1}{ s^{2}} \cos^2(\phi/2) + s^{2} \sin^2(\phi/2) \right )\quad , \quad
  \langle \Delta p\rangle_{\tau\alpha}^2 =  \frac \xi 2\,  \left ( s^2 \cos^2(\phi/2) +\frac{ 1}{ s^{2}}  \sin^2(\phi/2) \right ) \, .
\end{equation}
The uncertainty statement appropriate for this context is the
Schr\"odinger--Robertson inequality:
\begin{equation}\label{SR}
  \langle \Delta q\rangle^2\,\langle \Delta p\rangle^2  \geq \frac {\xi^2} 4
   + \langle \widehat{F}\rangle^2 \,, \qquad
  \widehat{F} \equiv \{\q - \langle \q \rangle,\p - \langle \p \rangle \}_\text{sym} \,,
\end{equation}
with the anti-commutator $\{X,Y\}_\text{sym} \equiv XY+YX$. Employing (\ref{B1}) to evaluate $\widehat{F}$ gives
 $\langle \widehat{F}\rangle_{\tau\alpha} =  (\xi/4)\,( s^{-2}-s^2)\sin\phi $.

\begin{figure}
\includegraphics[width=4.5cm]{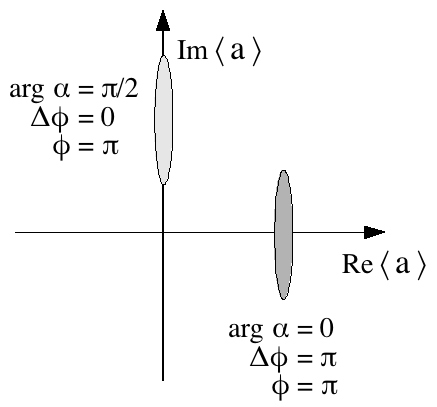}
\caption{\label{fig:squeezScheme}
Representation of squeezed states in the plane of complex amplitudes.
Each ellipse is centered at the mean value $\langle \hat{a} \rangle $ with
the principal axes corresponding to the uncertainties of the canonical variables.
The length of the axes is determined by the squeezing parameter $s$ and the direction
of the ellipse by the phase factor $\phi$.
The parameter $\Delta \phi$ appears in Eq.(\ref{4J}).}
\end{figure}

  Combining these statements shows that the $\tau\alpha$ squeezed states are
   minimum uncertainty states  with respect to the Schr\"odinger--Robertson lower bound.  In fact \cite{Puri95} a state that fulfills the equality in (\ref{SR}) must be a squeezed state.
 We remark that the phase $\phi$ of the squeezing parameter $\tau$ determines which of the canonical variables
is squeezed: for $\phi=0$ ($\phi = \pi$) the variance of $p$  ($q$) is reduced
by a factor of $s$, respectively. In other words, the angle variable $\phi/2$
rotates the semi-axis of the uncertainty ellipse with respect to the $q,p$
axis. This behaviour is usually visualized by representing the squeezed state
as an ellipse in the complex $\alpha$-plane that indicates the uncertainties of
the canonical variables, see Fig.~\ref{fig:squeezScheme}.

\smallskip

The quantum phase space representation of the expectation value is the phase
space integral (\ref{IP2})
\begin{equation}\label{4H} \langle \a(t)\rangle_{\tau\alpha} =
    \langle \tau\alpha |\a(t)| \tau\alpha \rangle = \frac 1{2\pi \xi}
    \int   \Theta_{01}(t|x) \big[|\tau\alpha\rangle
    \langle\tau\alpha |\big]_{{\rm\scriptstyle w}}(x)\,d^{\,2}x
\end{equation}
where $\Theta_{01}(t|x)$ is the symbol of $\a(t) = U_t^\dag\, \a\, U_t$.
\smallskip

 Next we compute the Weyl symbol of $\big[|\tau\alpha\rangle
    \langle\tau\alpha |\big]_{{\rm\scriptstyle w}}$ by relating it to the simpler quantity $\big[|\alpha\rangle
    \langle\alpha |\big]_{{\rm\scriptstyle w}} $. From the coherent state wave function
\begin{equation}\label{4F}
\langle q |\alpha \rangle  = \left( \frac 1{\pi \xi } \right)^{1/4} \exp \left[
\frac 1{\xi }\Big( -\frac {q^2}2 + \sqrt {2}\, \alpha\,q -
\alpha\,\Re\,\alpha\Big) \right]\, ,
\end{equation}
one obtains the associated Wigner distribution. Let
$\overline{x}=(\overline{q},\overline{p}) = (\sqrt 2\, \Re\,\alpha, \sqrt 2\,
\Im\,\alpha)$ be the $\alpha$ coherent state mean values, then one has
\begin{equation}\label{coh1}
{ \big[ |\alpha\rangle   \langle\alpha |\big ]_{{\rm\scriptstyle w}} (q,p) = 2
}
     \exp \xi^{-1}\left\{ - (q^2 + \, p^2)
    +  2q\,\overline{q}  + 2p\,\overline{p} - (\overline{q}^2 + \overline{p}^2) \right\}\,.
\end{equation}  The Weyl symbol (\ref{coh1}) is real because
$|\alpha\rangle \langle\alpha |$ is hermitian. The squeezed generalization of
this follows from the Weyl symbol covariance property (\ref{CV})
\begin{equation}\label{sqdensity}\begin{split}
  \big[|\tau\alpha\rangle
    \langle\tau\alpha |\big]_{{\rm\scriptstyle w}}(x) &= \big[V(\tau) |\alpha\rangle\langle\alpha|
     V(\tau)^\dag\big]\wig(x) = \big[|\alpha\rangle\langle
     \alpha|\big]\wig\big(S(\tau)\,x\big)\\ \vphantom{\Bigg]}
   &= 2\exp \frac 1\xi\Big[ -x\cdot S(2\tau)\, x +2x\cdot S(\tau)\, \overline{x} -\overline{x}\cdot\overline{x}\,
   \Big]\,. \end{split}
\end{equation}
The density matrix $|\tau\alpha\rangle \langle\tau\alpha |$ is projection operator
that characterizes a pure ensemble of photons with mean number
\begin{equation*}\label{photons}
    \langle \a^\dag \a \rangle_{\tau\alpha} = \sinh^2(2\xi|\tau|) +\big |\alpha
    \cosh(2\xi|\tau|) +
    \alpha^* e^{i \phi} \sinh(2\xi|\tau|)\big|^{\,2}\,.
\end{equation*}
The photon number is a constant of motion since $\a^\dag \a$ commutes with the
Kerr Hamiltonian.

The integral (\ref{4H}) is conveniently computed by diagonalizing $S(\tau)$ and $S(2\tau)$. The phase space rotation
\begin{equation*}
    R(\phi) = \left(
                \begin{array}{cc}
                  \cos \phi/2 & -\sin \phi/2 \\
                  \sin \phi/2& \cos \phi/2 \\
                \end{array}
              \right)
  \,, \quad\quad
   R(\phi)^T  = R(\phi)^{-1} = R(-\phi)
\end{equation*}
achieves this via
\begin{equation*}
    S(\tau) = R(\phi)\,
  \Lambda(s)
    R(-\phi)
    \,, \quad \quad
  \Lambda(s)= \left(
                                                                \begin{array}{cc}
                                                                 s  & 0 \\
                                                                  0 & s^{-1} \\
                                                                \end{array}
                                                              \right)\,.
\end{equation*}
Note the $S(\tau)$ eigenvalues $\lambda_1 = s,\ \lambda_2 = s^{-1}$ are
independent of $\phi$ and likewise the matrix $R(\phi)$ is independent of $s$.
The eigenvalues are positive because $S(\tau) > 0$ and are mutual
inverses since $\det S(\tau) = 1$.

Introduce the variables $y=R(-\phi)\,x$ and
\begin{equation*}\label{Y1}
\overline{y} = R(-\phi)\,\overline{x} =\left(
                \begin{array}{c}
                  \overline{q}\,\cos(\phi/2) + \overline{p}\,\sin(\phi/2) \\
                  -\overline{q}\,\sin(\phi/2)+ \overline{p}\,\cos(\phi/2) \\
                \end{array}
              \right) = \left(
                          \begin{array}{c}
                            \overline{q}_\phi \\
                            \overline{p}_\phi \\
                          \end{array}
                        \right)
              \,.
\end{equation*}
Changing the integration variable from $x$ to $y$ and employing the identity
$\Theta_{01}(t|R(\phi)y) = \exp(i\phi/2)\, \Theta_{01}(t|y)$ allows one to
write the integral (\ref{4H}) as
\begin{equation}\label{4Hb}
    \langle \a(t)\rangle_{\tau\alpha} = \frac {e^{i\phi/2}} {\pi \xi}
    \int   \Theta_{01}(t|y)\, \exp \frac 1\xi \Big[ -y \cdot
    \Lambda(s^2)\,y +
    2y\cdot\Lambda(s)\,\overline{y} - \overline{y}^{\,2} \,\Big]  \,d^{\,2}y\,.
\end{equation}

In displaying the final result it is useful to use the abbreviations
\begin{equation*}\label{4H2}
T = \tan(\xi \omega_2 t)
  \,, \quad \quad
    G(T,s) =\frac {(1+iT)^{\,2}}{(1+ i\,s^{-2}\, T)(1+
i\,s^2\,T) }\,.
\end{equation*}
Integral (\ref{4Hb}) is a generalized Gaussian integral and evaluates to
\begin{equation} \label{4J}\begin{split}
  \langle \a(t)\rangle_{\tau\alpha} =& \alpha\,
   G^{\,3/2}
   \left [\frac{ 1}{s}\, \cos \Big(  \xi\omega_2 t - \frac{ \Delta \phi }{2} \Big )
  + i s\,  \sin \Big (  \xi\omega_2 t - \frac{ \Delta \phi }{2} \Big ) \right ] \exp\big(-i(\omega_1 t
   + \xi\omega_2 t -\Delta
  \phi/2) \big)
\\   &\times \exp\left\{-2i \frac{ T}{\xi}  \frac{|\alpha|^2 \, G}{(1+i T)^2}
   \Big [
    i T + s^{-2} \cos^2(\Delta \phi /2) + s^2  \sin^2(\Delta \phi /2)
   \Big]
  \right\}\,. \end{split}
\end{equation}
Here $\Delta \phi = \phi - 2\, \text{arg}(\alpha)$ is the difference between the squeezing angle and twice the
phase of the coherent state amplitude $\alpha$ \footnote{ The factor of two
appears because a phase shift $\delta$ in the operator $\hat{a} $ changes
$D(\alpha)$ to
   $D(\alpha e^{-i\delta})$ but $S(\tau)$ to $S(\tau e^{-2i\delta})$}.
The branch cut for $\sqrt G$ lies along the positive real axis.

The result above constructs the quantum mean $\langle \a(t)\rangle_{\tau\alpha}$ directly from the Moyal solution $\Theta_{01}(t|x)$. It describes in detail the dependence of the expectation value on the semiclassical scaling parameter $\xi$  as well has the squeezing and coherent state variables, $\tau$ and $\alpha$. The $\langle \q(t)\rangle_{\tau\alpha}$ and $\langle \p(t)\rangle_{\tau\alpha}$ predictions are obtained from the real and imaginary parts of $\langle \a(t)\rangle_{\tau\alpha}$. We remark that result (\ref{4J}) agrees with an alternative derivation that does not use phase space techniques but employs the su(1,1) group structure of squeezing operators instead.

Formula (\ref{4J}) demonstrates that the functional structure of $ \alpha^{-1}  \langle \hat{a(t)} \rangle_{\tau\alpha} $ with respect to four initial state parameters $\{\tau,\alpha\}$  depends on just the three variables $|\alpha|,s,\Delta \phi$. At $t=0$, the case $\Delta\phi =0$ corresponds to phase squeezing: the uncertainty of the magnitude of $\langle \a\rangle_{\tau\alpha}$ is increased and that of its phase factor is reduced. An example for phase squeezing is the lighter ellipse shown in Fig.~\ref{fig:squeezScheme}. On the other hand, $\Delta\phi =\pi$ corresponds to number squeezing: the uncertainty of $|\langle \a\rangle_{\tau\alpha}|$, which is the square root of the mean number of photons, is decreased. This is the case for the darker ellipse shown in Fig.~\ref{fig:squeezScheme}.

Note that solution $\Theta_{01}(t|x)$ has different frequencies for different $|x|$. Thus the minimum uncertainty character of the initial state $|\tau\alpha\rangle$ is lost for $t\neq 0$   and restored at half-period multiples, $\xi\omega_2 t = N \pi$.

An important feature of the expectation value (\ref{4J}) is that for fixed $s>0$ it is a smooth bounded function in all variables; in particular it does not display the Moyal solution singularity when $\xi \omega_2 t$ approaches $\pi/2$. In order to interpret this we recall how the Heisenberg uncertainty principle works in the quantum phase formalism. Weyl symbols are often distributions and as such do not have any restrictions on localization or magnitude.  For example, the quantizer $\widehat\Delta(x')$ \cf(\ref{A2}) is a bounded operator whose symbol is the delta function $\delta(x'-x)$. The information about phase space uncertainty is encoded in the Wigner function $\big[|\tau\alpha\rangle\langle\tau\alpha |\big]_{\rm\scriptstyle w}(x)$. Only when the expectation value integral (29) is evaluated are the full effects of quantum uncertainty imposed. This phase space integration averages out the Moyal solution singularity giving a finite result.

\begin{figure}
a)
\includegraphics[width=7cm]{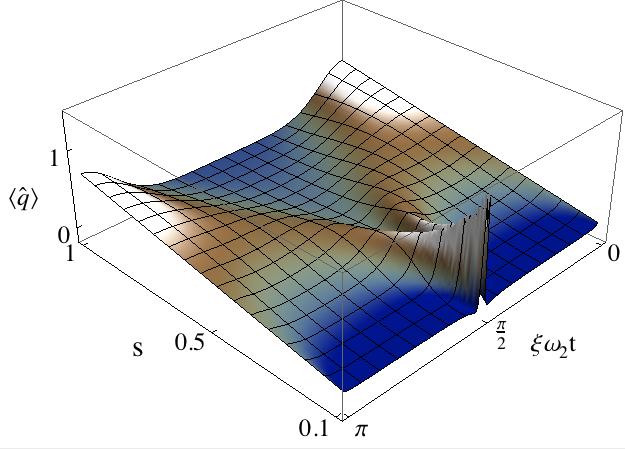}
\hspace{1cm}
b)
\includegraphics[width=7cm]{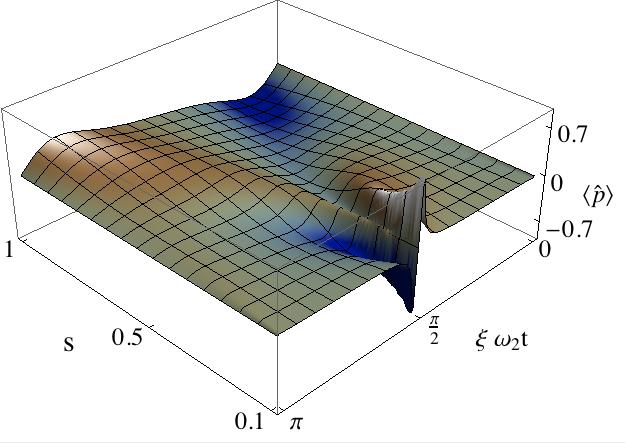}
\caption{\label{fig:numSqueeze}
Mean a) position and b) momentum for number squeezed states ($\Delta\phi=\pi$)
for the case $\alpha = \xi =1$.}
\end{figure}

\begin{figure}
a)
\includegraphics[width=7cm]{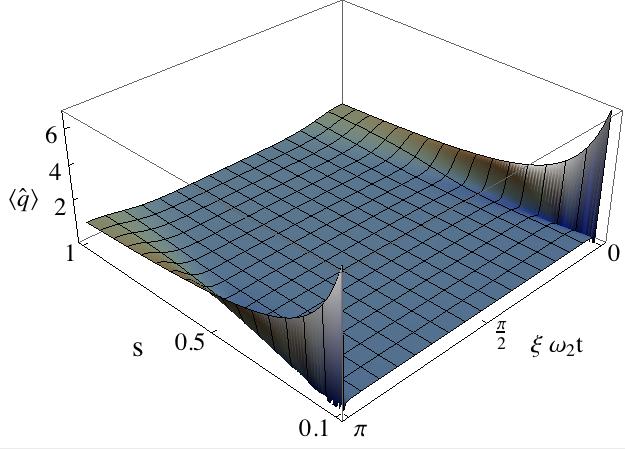}
\hspace{1cm}
b)
\includegraphics[width=7cm]{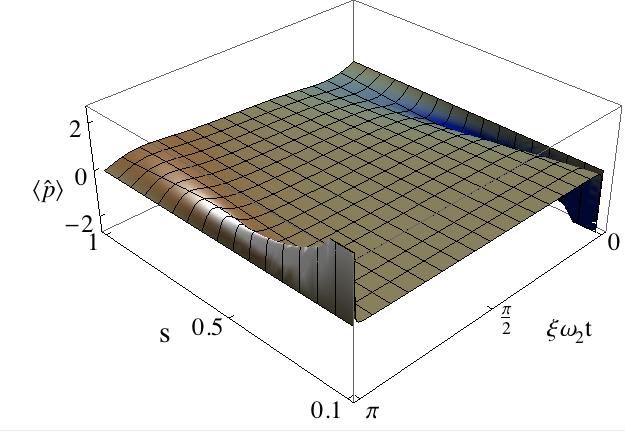}
\caption{\label{fig:phaseSqueeze}
Mean a) position and b) momentum for phase squeezed states ($\Delta\phi=0$)
for the case $\alpha = \xi =1$.}
\end{figure}

However, one may ask whether there is a surviving signature of the Moyal singularity at $\xi \omega_2 t =\pi/2$ in the observables $\langle\q(t)\rangle_{\tau\alpha}$ or  $\langle \p(t)\rangle_{\tau\alpha}.$ For suitable values of $\tau,\alpha$ the answer is yes.
In the next section
we will present a general argument that these effects should be most evident for
states similar to number-squeezed states, $\Delta\phi =\pi$. In
Fig.~\ref{fig:numSqueeze}  it can be seen that this is indeed the case: for strong number squeezing ($s\ll 1,\Delta\phi =\pi$)
position and momentum mean values are significantly enhanced around the singular point On the other hand, in the case of
phase squeezing ($\Delta\phi =0$) the singular behavior of the Moyal solution has virtually no effect on the mean amplitude
(Fig.~\ref{fig:phaseSqueeze}). We remark that the peak in Fig.~\ref{fig:phaseSqueeze} appears at $t=0$ and therefore
corresponds to the mean value of position and momentum in a squeezed coherent state in absence of the nonlinear
interaction.

Fig.~\ref{fig:numSqueeze} a) suggests that a singularity does appear
in the limit of infinite number squeezing. However, for any fixed value $\xi
\omega_2 t \neq \pi/2$ we have $\lim_{s\rightarrow 0} \langle \hat{a}(t)
\rangle_{\tau\alpha} =0$. On the other hand, for fixed $s$ we have
$$
  \lim_{\xi \omega_2 t \rightarrow \pi/2}\langle \hat{a}(t) \rangle_{\tau\alpha}  = \frac{ 1}{s}
  \exp \left ( -2 \frac{ |\alpha|^2}{\xi} \right )\; .
$$
This indicates that the peak becomes infinitely narrow in the limit of infinite squeezing and thus is not of physical significance. We remark that in current experiments a squeezing factor of about $s=0.1$ can be achieved \cite{vahlbruch:033602}.

The general expectation value simplifies dramatically in several special cases. For purely coherent states, $\tau=0$,  expression (\ref{4J}) reduces to the well-known result (see, e.g.,  Eq.~(40) of Ref.~\cite{shapiro:josab1993})
\begin{equation}\label{nosqueeze}
    \langle \a(t)\rangle_{\tau\alpha}|_{\tau=0}\ = \alpha\exp\left\{-i\omega_1t
     - 2i \frac {|\alpha|^2}{\xi} \sin(\xi\omega_2 t)\, e^{-i(\xi\omega_2 t)} \right\}\,.
\end{equation} This no-squeezing result does not show any evidence of the Moyal solution singulary.

Next consider the small $\xi$ behavior of $\langle \a(t)\rangle_{\tau\alpha}$. In this limit all the noncommuting effects of the Heisenberg algebra for $\q,\p$ are turned off. The Weyl symbol based semiclassical expansion is constructed by replacing $\Theta_{01}$ with its semiclassical approximation (\ref{SC1}) in the evaluation of the integral (\ref{4Hb}). One finds
\begin{equation}\label{4L}
\langle \a(t)\rangle_{\tau\alpha} =\, a_\text{cl}(t|\overline{x})\, \bigg\{1 + \Big[2|\tau| e^{\,i\Delta\Phi}  - 2|\alpha|^2 \omega_2 t \Big(\omega_2 t + 4i |\tau| \cos(\Delta\Phi) \Big) \Big]\xi + \O(\xi^2)  \bigg\}\,.
\end{equation}
The leading $\xi=0$ term term is just the complex statement of the Kerr classical flow (\ref{3Fa}) with the initial condition $(q,p) = (\overline{q}, \overline{p}) ={\sqrt 2}\,(\Re\,\alpha,\Im\,\alpha)$.  The formula (\ref{4L}) may also be obtained by implementing a power series expansion of (\ref{4J}).  However the procedure using $\Theta_{01}^{\text{sc}}(t|x)$ has wider application in that it does not require an exact Moyal solution. The $\xi$-linear correction factor senses the dependence on the squeezing $|\tau|$ and the phase $\Delta\Phi$. For this expansion to be a good approximation to $\langle \a(t)\rangle_{\tau\alpha}$ the factor $1$ in the curly bracket must be much larger than the correction terms. For $|\alpha|=1$ this requires $s \in (0.8,1.0)$ and $|\omega_2 t | \ll 1$. Though this region of good approximation is an extremely small portion of the variable range shown in Figs.~\ref{fig:numSqueeze} and \ref{fig:phaseSqueeze}, it nevertheless covers a significant part of the experimentally accessible range.

\section{Finite Expectation Values}\label{sec-expval2}
The result that the exact solution for the Weyl symbol of a well-established physical observable diverges periodically in time is surprising. Such time-periodic singularities are not present in the harmonic oscillator basis or in the classical solution for the Kerr Hamiltonian. In this section we therefore examine for general initial states whether such singularities could in principle survive the phase space averaging integral that defines an expectation value in the Weyl symbol picture. The results of the previous section show that the squeezed state expectation values of $\Theta_{01}(t|x)$ are finite for all times.  Here we show that the expectation values of the general solution $\Theta_{sm}(t|x)$, whose amplitude diverges as $(\sec \tilde{t})^{s+m+1}$, are finite for almost all possible quantum states.

A very useful property of coherent states is their over-completeness,
\begin{equation*}
  I = \frac{ 1}{\pi \xi} \int |\alpha \rangle \langle \alpha
  |\, d^2\alpha \,.
\end{equation*}
We therefore can express the expectation value of an operator $\hat{f}$ as
\begin{equation*}
  \text{Tr}(\rho \hat{f}) = \frac{1 }{(\pi\xi)^2}
  \int   \langle \beta|\rho
  |\alpha \rangle
  \langle \alpha |\hat{f}|\beta \rangle\, d^{\,2}\alpha \, d^{\,2}\beta \; ,
\end{equation*}
\noindent with  $\rho$ the density matrix of the initial state of the system.
Hence, to see if the singularities can appear for any quantum state it is
sufficient to investigate the matrix element $ \langle \alpha |\hat{f}|\beta
\rangle$ of an operator. Using Eq.~(\ref{A3}) we can express this matrix
element as
\begin{equation*}
  \langle \alpha |\hat{f}|\beta \rangle =
  \int f(x)\, \langle \alpha |\widehat{\Delta}(x)|\beta \rangle\, d^{\,2}x \,,
\label{eq-meanf} \end{equation*}
with $f$ the Weyl symbol of $\hat{f}$. In the
following we will evaluate the integral over
$d^{\,2}x$ for the
set of Weyl symbols (\ref{MA1}). Using Eqs.~(\ref{4F}) and (\ref{eq-qDq}) it is
not hard to see that, in complex coordinates $z=q+ip$,
\begin{eqnarray}
  \langle \alpha |\widehat{\Delta}(x)|\beta \rangle &=&
   \frac{ 1}{\pi\xi} \exp \left (
     -\frac{z z^*}{\xi} + \frac{ \sqrt{2}}{\xi}(\beta z^* + \alpha^*
     z) +C \right )
\label{aDb}
\end{eqnarray} where $ C \equiv  -
    [ |\alpha|^2 + |\beta|^2  + 2\beta \alpha^*]/(2\xi).$\smallskip

 Using Eqs.~(\ref{MA1}),(\ref{aDb}), $d^{\,2}x={\frac 1
2}dz\,dz^*$ and (\ref{IP2}) this leads to
\begin{eqnarray*}
 &&\langle \alpha | (\hat{a}(t)^\dagger)^s\, \hat{a}(t)^m | \beta \rangle
 =
  \int\Theta_{sm}(t|x) \, \langle \alpha
  |\widehat{\Delta}(x)|\beta \rangle\,  d^{\,2}x
\nonumber \\ &&\qquad=
    e^{-i(m-s) \omega_1 t}(\sec \tilde{t})^{m+1}  e^{ i(2- s) \tilde{t}}
   \, \frac 12 \int
  \langle \alpha |\widehat{\Delta}(x)|\beta \rangle
  \left ( \frac{ z^* - \xi \partial_z}{\sqrt{2}} \right )^s
  \exp \left (
  -\frac{i}{\xi} z z^* \tan (\tilde{t})
  \right )
   \left ( \frac{ z}{\sqrt{2}}\right )^m {dz\, dz^*}\,
\nonumber \\ &&\qquad=  e^{-i(m-s) \omega_1 t}
   (\sec \tilde{t})^{m+1} e^{ i (2- s) \tilde{t}}
   \left (\alpha^* \right )^s \, \frac 12 \int
  \langle \alpha |\widehat{\Delta}(x)|\beta \rangle
  \exp \left (
  -\frac{i}{\xi} z z^* \tan (\tilde{t})
  \right )
   \left ( \frac{ z}{\sqrt{2}}\right )^m {dz\, dz^*} ,
\end{eqnarray*}
where we have performed a partial integration and used that $ ( z^* + \xi
\partial_z)
  \langle \alpha |\widehat{\Delta}(x)|\beta \rangle
= \sqrt{2} \alpha^*  \langle \alpha |\widehat{\Delta}(x)|\beta \rangle $.
Performing the integration yields
\begin{eqnarray*}
  \langle \alpha | (\hat{a}(t)^\dagger)^s\, \hat{a}(t)^m | \beta \rangle
  &=&
   \frac{ \left (\alpha^* \right )^s }{2\pi\xi} (\sec \tilde{t})^{m+1} e^{-i(m-s)
    \omega_1 t} e^{C+ i (2- s) \tilde{t}}
   \left ( \frac{ \xi}{2} \partial_{\alpha^*}
  \right )^{m}
  \int
  e^{   -\frac{z z^*}{\xi} (1+i \tan\tilde{t}) + \frac{ \sqrt{2}}{\xi}(\beta z^* + \alpha^*
     z)
  } {dz\, dz^*}
\nonumber \\ &=&
  \left (\alpha^* \right )^s  \beta^m  e^{-i(m-s)t (\omega_1+ (m+s-1)\xi \omega_2) }
  \exp \left (
     - \frac{ 1}{2\xi}
    \left ( |\alpha|^2 + |\beta|^2\right )  + \frac{  \beta
    \alpha^*}{\xi} e^{-2i (m-s)\xi \omega_2 t}
  \right )\,.
\end{eqnarray*}

This result is in perfect agreement with the corresponding expression derived
from the solution of the ordinary Heisenberg equations of motion and thus
demonstrates the consistency of the Moyal-Kerr solution (\ref{MA2}). For our
discussion it is important to note that none of these matrix elements contains
a singularity. This means that the singularity can be considered as a feature
of the Moyal representation of quantum mechanics that does not generate
divergent expectation values.

An intuitive explanation of why the singularity does not show up in expectation
values is as follows. At the time $\tilde{t}$ when the amplitude of
Eq.~(\ref{MA2}) diverges, the phase factor diverges as well. However, the
diverging factor in the phase contains  the mean photon number $z z^*$.
Consequently, the phase divergence is different for states with different
photon numbers. Therefore, for $\tilde{t}$ sufficiently close to the singular
value $\pi/2$, the phase factor would average to zero for any state that has a
variance in the number of photons.

Hence, the only chance to see the singularity would be in a state where the
number of photons is exactly, known, i.e., number states. But number states
correspond to states for which the phase is completely undetermined, so that
any expectation value with $s\neq m$ would average to zero for any value of
$\tilde{t}$. Hence, the uncertainty relation $\Delta n \, \Delta\phi >1/2$
(see, e.g., Ref.~\cite{wolf}) for photon number and phase prohibits the appearance of the Moyal singularity in a quantum photon observation (\ref{MA2}).

\section{Conclusion}
In this paper we have derived an exact solution (\ref{MA1}) for the Weyl symbol phase space representation of the Kerr model of nonlinear quantum optics. This solution exhibits singularities that are absent for the classical solution and depend on the intensity of the light field. This intensity-dependence guarantees that expectation values for all initial states remain finite. On the other hand, a signature of the singularity appears in the form of finite peaks in the expectation values (\ref{4J}) for number-squeezed states of light. A number of open questions remain. The single-mode Kerr model that we have studied gives only a good description for photons in optical cavities of extremely high finesse. A generalization of our results for a multi-mode theory of propagating photons would therefore be desirable. Alternatively, an imperfect cavity could be modeled by studying a Kerr model that is coupled to the environment and exhibits Langevin noise. Both of these aspects have been addressed by K\"artner {\em et al.} \cite{kartner:QOpt1992} in the context of a specific noise model using the Wigner function. The corresponding Weyl symbol representation of the Heisenberg-Langevin equations of motion may be free of divergences and would open new perspectives for phase space descriptions of quantum systems.

\acknowledgments
K.-P.~M. wishes to thank Barry Sanders and Alex Lvovsky for helpful
discussions. The authors thank Frank Molzahn for a critical reading of the text. Financial support by NSERC and iCORE is gratefully acknowledged.
\begin{appendix}


\section{Weyl symbol quantum mechanics}\label{AppendixWeyl}
This appendix summarizes properties of the quantum phase space method that are employed in this paper. We collect the various Weyl symbol identities in a notation suitable for quantum optics. The account below closely matches that found in Refs.~\cite{OM95,KO3}.

For a single-mode photon and a suitable choice of reference point, the electric field strength $E$ is proportional to $\a + \hat{a}^\dagger $ (or equivalently to $\q$) and the magnetic field strength $B$ to $i(\a - \hat{a}^\dagger )$ or $\p$. For this reason the phase space base manifold for a single mode state is the real line $\R$. The noncommutivity of $\q$ and $\p$ arise from the mode operators $\a$ and $\hat{a}^\dagger $.  The quantum state space is that spanned by the harmonic oscillator basis, or equivalently the Hilbert space of one dimensional square integrable wave functions, $\H = L^2(\R,\C)$. Likewise, the associated classical phase space is $\,T^*\R = \R^2$ equipped with the standard Poisson bracket.

Weyl quantization maps functions on $\,T^*\R$ into operators on $\H$. A unified characterization of both quantization and de-quantization is achieved via a \textit{quantizer} \cite{STR57,Roy77,Gro76}. Let $\{\widehat{\Delta}(x): x=(q,p) \in T^*\R\}$ be a $x$-dependent family of bounded, self-adjoint operators on $\H$ defined by their action on a wave function, $\psi$
\begin{equation}\label{A2}
   \psi'(q')  = [\widehat{\Delta}(q,p) \psi](q') \equiv \frac{1}{\pi\xi}\exp\left( {\frac{2i}{\xi}
    p\,(q'-q)}\right)
    \psi(2q-q')\,,
\end{equation}
or, equivalently, as an integral kernel
\begin{equation}
  \langle q' | \widehat{\Delta}(x)| q'' \rangle =
  \frac{ 1}{\pi\xi} \exp \left (
    \frac{ i}{\xi} p(q'-q'')
  \right ) \delta (2q- q'-q'') \; .
\label{eq-qDq}\end{equation}

Then both quantization and dequantization are constructed from
$\widehat{\Delta}(q,p)$ via
\begin{equation}\label{A3}
    \f = \int_{T^*\R} f(x)\,\widehat{\Delta}(x)\, d^{\,2}x\,,  \quad\qquad [\f]_{{\rm\scriptstyle w}}(x) =
    (2\pi\xi)\,
    \text{Tr}\,\f\,\widehat{\Delta}(x)
\end{equation}
where $\text{Tr}$ is the trace on $\H$. This pair of linear transformations are
mutual inverses, so $f=[\f]_{{\rm\scriptstyle w}}$. The notation
$[\f]_{{\rm\scriptstyle w}}$ indicates the Weyl symbol of the operator $\f$.
The second identity in Eq.~(\ref{A3}) is proportional to the Wigner transform of $\f$, \cf
(\ref{eq:wignerfunction}).

This bijective correspondence between phase space functions and operators is
simple in a variety of important cases. For example, operators $f(\q)$, $g(\p)$
and the identity on $\H$ have symbols $f(q)$, $g(p)$ and the constant function
$1$. The quantizer has symbol, $[\widehat{\Delta}(x)]\wig(x') = \delta(x-x')$; in
turn this implies that the exponential operator $e^{i u\cdot \hat x}, \,\, u\in \R^2$ has the symbol $e^{i u\cdot x}$.

The Weyl symbol framework is a Hilbert algebra, namely a complete linear space
$L$ with three basic structures: an associative product $\star$, an involution
$^*$ and an inner product $(\cdot,\cdot)_{L}$.  The product of operators on
$\H$ is mirrored by the noncommutative product of Weyl symbols. This star
product is defined by $f \star g \equiv [\f\,\g]_{{\rm\scriptstyle w}}$. Given
the $\star$ product, the Moyal bracket is defined as
\begin{equation}\label{Moyal1}
    \{f,g\}_M = \frac 1{i\xi} [\f,\g]\wig = \frac 1{i\xi}\big(f\star g - g\star
    f\big)\,.
\end{equation}

The $\star$ product has three useful representations.\smallskip

The first is Berezin's integral form \cite{FAB72}
\begin{equation}\label{Ber1}
f \star g(x) = \frac1{(\pi\xi)^2}\int\!\!\int\ f(x_1)\,g(x_2)\exp \left\{
\frac{2i}{\xi} (x_1\wedge x_2 + x_2 \wedge x + x \wedge
x_1)\right\}\,d^{\,2}x_1\, d^{\,2}x_2\,.
\end{equation}  Here
$x_1\wedge x_2 \equiv x_1\cdot J x_2$.

Next is Groenewold's derivative expansion  \cite{Gro46}
\begin{equation}
f \star g(x) =  \exp\left( \frac{i\xi}{2}\partial_1\!\cdot\! J
\partial_2 \right) \prec\! f,g \succ(x)\,.
\label{Gro1}\end{equation} Above $\partial_1$ and $\partial_2$ are gradients
acting on the first ($f$) and second ($g$) arguments of the product $\prec\!
f,g \succ$. The Poisson bracket $\{f,g\}$, in this notation, is
$\partial_1\!\cdot\! J \partial_2  \!\prec\! f,g \succ\, $ followed by diagonal
evaluation, $x_1=x_2=x$. In the case where $f,g$ are $\xi$ independent and
suitably smooth, the series expansion of Eq.~(\ref{Gro1}) gives a
$\xi$-asymptotic expansion of $f \star g$ with the Poisson bracket term as the
leading semiclassical correction.

The third realization of the $\star$ product is the left, right form. Define
\begin{equation}\label{LR1}
    \L \equiv x + \frac {i\xi}{2} J \partial_x \qquad
    \RR \equiv x - \frac {i\xi}{2} J
    \partial_x\,,
\end{equation} then for smooth $f,g$
\begin{equation}
    f \star g\,(x) = \big(f(\L)\,g)\,(x) = \big(g(\RR)\,f)\,(x)\,.
\label{LR2} \end{equation} The differential operators $\mathcal L$ and $\mathcal R$ commute.

The involution operation on $L$ is complex conjugation.  It is the symbol
analog of the adjoint operation on $\H$. If $f=[\f]_{{\rm\scriptstyle w}}$
 then $f^* = [\f^\dag]_{{\rm\scriptstyle w}}$.

For suitably restricted $\f,\g$ (\eg both Hilbert--Schmidt), the trace of the
operator product defines the $L$ inner product, in detail
\begin{equation}\label{IP2}
  (2\pi\xi)\,\text{Tr}\, \f\,\g = \int_{\R^2} f\!\star g\,(x) \,d^{\,2}x
    = \int_{\R^2} {f(x)}\,g(x) \,d^{\,2}x  = (f^*,g)_{L}\,.
\end{equation}
This formula shows that quantum expectation values in $L$ are
obtained by phase space integration. For example, this occurs if $\f$ is a
density matrix and $\g$ is any observable.

Weyl quantization has a simple covariance property. A unitary operator $V$ is called
{\it{metaplectic}} if $V \x\, V^\dag = S \x$ for some symplectic matrix $S$, i.e. $SJS^T = J$.
If  $\f$ has the symbol $f$, the affine canonical covariance property \cite{OM95}
is the statement that
\begin{equation}\label{CV}
[V \f\, V^\dag]\wig(x) = f(S\,x)\,.
\end{equation}

\end{appendix}

\bibliography{MoyalKerr18a}

\end{document}